\documentclass[12pt,preprint]{aastex}

\shorttitle{Foreground contribution to the BEAST CMB maps}
\shortauthors{Mej\'{\i}a et al.}

\begin{document}
\def\ts{\thinspace}

\title{Galactic foreground contribution to the BEAST CMB
Anisotropy Maps}

\author{Jorge Mej\'{\i}a \altaffilmark{1}
Marco Bersanelli \altaffilmark{2}
Carlo Burigana \altaffilmark{3}
Jeff Childers \altaffilmark{4}
Newton Figueiredo \altaffilmark{5}
Miikka Kangas \altaffilmark{4}
Philip Lubin \altaffilmark{4}
Davide Maino \altaffilmark{2}
Nazzareno Mandolesi \altaffilmark{3}
Josh Marvil \altaffilmark{3}
Peter Meinhold \altaffilmark{4,6}
Ian O'Dwyer \altaffilmark{7}
Hugh O'Neill \altaffilmark{3}
Paola Platania \altaffilmark{2}
Michael Seiffert \altaffilmark{8}
Nathan Stebor \altaffilmark{3}
Camilo Tello \altaffilmark{1}
Thyrso Villela \altaffilmark{1}
Benjamin Wandelt \altaffilmark{7}
Carlos Alexandre Wuensche \altaffilmark{1}
}

% Affiliations
\altaffiltext{1}{Instituto Nacional de Pesquisas Espaciais, Divis\~ao de Astrof\'{\i}sica, Caixa Postal 515, 12210-070 - S\~ao Jos\'e dos Campos, SP, Brazil}

\altaffiltext{2}{Dipartimento di Fisica, Università degli study di Milano, via Celoria 16, 20133 Milano, Italy}

\altaffiltext{3}{IASF-CNR sezione di Bologna, via P.Gobetti, 101, 40129 Bologna, Italy}

\altaffiltext{4}{Physics Department, University of California, Santa Barbara, CA 93106}

\altaffiltext{5}{Universidade Federal de Itajub\'a, Departamento de F\'{\i}sica e Qu\'{\i}mica, Caixa Postal 50, 37500-903 - Itajub\'a, MG, Brazil}

\altaffiltext{6}{University of California, White Mountain Research Station, CA 93514}

\altaffiltext{7}{Astronomy Department, University of Illinois at Urbana-Champaign, IL 61801-3074}

\altaffiltext{8}{Jet Propulsion Laboratory, California Institute of Technology, Oak Grove Drive, Pasadena, CA 91109}

%Corresponding author
\email{mejia@das.inpe.br}

\begin{abstract}
We report limits on the Galactic foreground emission contribution
to the Background Emission Anisotropy Scanning Telescope (BEAST)
Ka- and Q-band CMB anisotropy maps. We estimate the contribution from
the cross-correlations between these maps and the foreground emission
templates of an H${\alpha}$ map, a de-striped
version of the Haslam et al. 408 MHz map, and a combined
100 $\mu$m IRAS/DIRBE map. Our analysis samples the BEAST
$\sim10^\circ$ declination band into 24 one-hour (RA) wide sectors with
$\sim7900$ pixels each, where we calculate: (a) the linear correlation
coefficient between the anisotropy maps and the templates; (b) the
coupling constants between the specific intensity units of the templates
and the antenna temperature at the BEAST frequencies and (c) the
individual foreground contributions to the BEAST anisotropy maps.
The peak sector contributions of the contaminants in the Ka-band are of 56.5\% 
free-free with a coupling constant of $8.3\pm0.4$ $\mu$K/R, and 67.4\% dust 
with $45.0\pm2.0$ $\mu$K/(MJy/sr). In the Q-band the corresponding values 
are of 64.4\% free-free with $4.1\pm0.2$ $\mu$K/R and 67.5\% dust with
$24.0\pm1.0$ $\mu$K/(MJy/sr). Using a lower limit of 10\% in the relative
uncertainty of the coupling constants, we can constrain the sector
contributions of each contaminant in both maps to $< 20$\% in 21
(free-free), 19 (dust) and 22 (synchrotron) sectors. At this level, all
these sectors are found outside of the $\mid$b$\mid = 14.6^\circ$ region.
By performing the same correlation analysis as a function of Galactic scale
height, we conclude that the region within $b=\pm17.5^{\circ}$ should be
removed from the BEAST maps for CMB studies in order to keep individual
Galactic contributions below $\sim 1$\% of the map's rms.
\end{abstract}


\section{INTRODUCTION}
The study of the anisotropies in the Cosmic Microwave Background Radiation
(CMB) angular distribution represents one of the most
important and active areas in Cosmology today. Measurements of the
CMB anisotropies provide an important probe to understand the
mechanisms of structure formation in the early Universe.
Unfortunately, these measurements are hampered by Galactic and
extragalactic emissions which limit the accuracy of the measured
CMB power spectrum. As for Galactic contribution, diffuse emission
is mainly due to synchrotron, free-free (thermal Bremsstrahlung)
and dust emissions. Thus, it is important
to quantify the precise level of these contaminating foregrounds in order
to distinguish them from the cosmological signal. The clear identification
of the contribution of each individual contaminant of the CMB signal
is a challenging astrophysical task.

Below $\sim 50$ GHz, the primary Galactic foreground contaminants
are synchrotron and thermal Bremsstrahlung emission. Their spectral signature,
$T(\nu) \approx \nu^{-\beta}$, differ significantly from that of CMB
fluctuations, since $\beta_{synch} \simeq 2.7$ and $\beta_{ff} \simeq 2.1$.
Therefore, multi-frequency measurements with a large enough signal-to-noise
ratio can distinguish between foreground and CMB fluctuations. Above
$\sim 50$ GHz, the primary contaminant is interstellar dust emission, whose
spectral shape is well fitted by an expression of the type
$I_\nu \propto \nu^\beta B_\nu(T_d)$. At these frequencies, dust emission
can also be distinguished from CMB fluctuations by its distinct spectral
signature ($\beta_{dust} \simeq 1.5 - 2.0$) in multi-frequency observations.
Recently, an additional component, correlated with dust emission, 
has been proposed (Draine \& Lazarian 1998) but its existence and nature is 
still matter of discussion.
Since spatial variability of $\beta_{synch}$ and $\beta_{dust}$ is quite
significant, cross-correlation techniques between CMB maps and
Galactic foreground templates have been employed recently in order
to estimate  the contribution of Galactic foregrounds to CMB
anisotropy data sets (e.g. Banday et al. 2003; Bennett et al. 2003; 
Hamilton \& Ganga 2001; de Oliveira-Costa et al. 1999).

In this paper we evaluate the Galactic emission contribution to
the BEAST (Background Emission Anisotropy Scanning Telescope)
CMB anisotropy maps at 30 GHz and 41.5 GHz (Meinhold et al. 2003).
We use a cross-correlation approach between the $\simeq 10^\circ$ wide
declination band observed by BEAST in the Northern Hemisphere
and the corresponding sky coverage in: (a) the Finkbeiner et al. (2003)
all-sky map of H$\alpha$ as a template for the free-free emission;
(b) the Platania et al. (2003) de-striped version of the 408 MHz map
(Haslam et al. 1982) as a template for the Galactic synchrotron emission and
(c) a 100 $\mu$m combined
IRAS/DIRBE dust template (Schlegel, Finkbeiner \& Davis, 1998) for the
dust emission. In the case of the free-free contribution, we compare our
results with those obtained by Dickinson et al. (2003). In Section 2 we
briefly describe the BEAST experiment. Section 3 presents the BEAST
maps and discusses the foreground templates. We describe the method
for evaluating the Galactic contamination in the BEAST maps in Section 4,
while Section 5 discusses the analysis and results of this work.

\section{THE BEAST TELESCOPE}
The BEAST instrument (Childers et al. 2004; Figueiredo et al.
2004) is a 1.9-meter effective aperture off-axis Gregorian telescope
configured with a 6-element Q-band (38-45 GHz; $23\pm 1$ arcmin FWHM) and 2-element
Ka-band (26-36 GHz; $28\pm 2$ arcmin FWHM) focal plane array, and a modulating tilted
flat mirror between the primary mirror and the sky. The instrument
is currently acquiring data in Barcroft, CA, USA, at the White
Mountain Research Station (WMRS) of the University of California at an
altitude of $\sim3800$ m. The spatial modulation provided by the
movement of the tilted flat mirror results in an ellipsoidal scan
of the sky with an $\sim10^\circ$ major axis. Combined with the daily
modulation due to the rotation of the Earth, BEAST scans a full 24-hour
long declination band between $+33\deg$ and $+42\deg$.

BEAST was designed to map the CMB with large sky coverage and
high angular resolution. It was specifically conceived to make well
connected maps through its scanning strategy. The range of frequencies
covered by BEAST can help discriminate between the CMB signal and
the Galactic foregrounds. The BEAST maps complement the Wilkinson
Microwave Anisotropy Probe (WMAP) mission in several ways.
For instance, BEAST measurements in the Q-band are of  higher
resolution than in WMAP, so we should be more sensitive to point
sources. By measuring the CMB with high sensitivity and resolution
over a limited region of the sky, we can get a better understanding of
foregrounds to help disentangle the individual Galactic contaminants.
BEAST can also be adapted to allow fundamental studies of the S-Z
effect on a large number of clusters, study dusty galaxies and their
foreground contaminating role, and extend its multipole coverage.

\section{BEAST AND FOREGROUND MAPS}
The data used in this work correspond to $\sim$680 hours of observations
taken between July 2001 and October 2002. An in-depth discussion of
the observational strategy, the data processing procedure, and the
map-generating pipeline is presented in Meinhold et al. (2003).
During the acquisition process, the amplified output of each radiometer
was AC-coupled with a high-pass time constant of 15 seconds. The raw
data were binned into 250 sky positions per revolution of the modulating
flat mirror and the first-order atmospheric contribution was eliminated
by folding and subtracting the low frequency envelope in one-hour long
data sets. A 10-Hz high-pass filter was applied to the resulting time-ordered
data (TOD) in order to remove any remaining low frequency contribution
of the $1/f$ noise, which is dominant in this region of the spectrum. The
pointing coordinates were translated into HEALPix (G\'orski et al. 1999)
pixel indexes to constitute the final BEAST maps per hour and per channel.
These individual maps were co-added to obtain the final BEAST maps
at 30 GHz and 41.5 GHz. Each BEAST map consists of
$\sim2 \times 10^{5}$ pixels (in HEALPix N$_{side}=512$
pixelization) covering the entire declination band between
$33^\circ<\delta<42^\circ$. In this work, the region limited by 
$\mid$b$\mid = 10^\circ$ was removed considering that its complexity
avoids any reasonable analysis. The resulting Ka-band map is shown in Figure 1
and the Q-band map in Figure 2.

The Galactic emission model we used for tracing the free-free component
is the recently released compilation of Finkbeiner (2003) of an all-sky map
of H$\alpha$ with observations from WHAM (Reynolds et
al. 2002), VTSS (Dennison et al. 1998) and SHASSA (Gaustad et al.
2001). As a tracer of Galactic synchrotron emission we used
the 408 MHz map of Haslam et al. (1982). Finally, to model
the thermal dust emission of the Galaxy we used a 100 $\mu$m combined
IRAS/DIRBE map (Schlegel, Finkbeiner \& Davis 1998; Finkbeiner,
Davis \& Schlegel 1999).

The three foreground maps were binned to the N$_{side}=512$ HEALPix
pixelization resolution and, then, processed by a BEAST experiment
simulator, which samples, individually, each foreground map
following the sequence of pixels observed by the BEAST
telescope during a typical working day, in 1-hour time sets in the
same manner the raw data are recorded.
This procedure effectively reproduces the BEAST scanning pattern of the sky
onto the foreground maps as if each Galactic component were to be the
only source in the sky. The resulting TODs were then processed following
exactly the same map-making pipeline as with the BEAST raw data, described
in the previous paragraph, to produce
BEAST-like templates of the Galactic emission tracers.

The pipeline to produce BEAST-like foreground templates can be summarized as follows:

\begin{itemize}

\item selection of the appropriate templates,

\item convolution of the original templates with the proper smooth function,

\item appropriate binning and processing of the individual templates according to BEAST pointing,

\item subtraction of the low frequency envelope and 10-Hz highpass filtering of the 1-hour long ``template TOD'',

\item processing of the ``template TOD'' into a foreground template map in the same way as described in Meinhold et al. (2003).

\end{itemize}

\section{METHOD}
In order to analyze the foreground contribution to the BEAST CMB
anisotropy maps, we considered a BEAST map, $T_{BEAST}$, to be a
linear superposition of the actual CMB anisotropy distribution,
$T_{CMB}$, a noise pattern, $n$, and a set of foreground
components $X_{i}$:

\begin{equation}
T_{BEAST} = T_{CMB} + n + \sum_i \alpha_i X_i,
\end{equation}

\noindent where $\alpha_i$ are the coupling constants which transform
the foreground template intensities into antenna temperature at
the BEAST frequencies. These constants correspond to the correlation
coefficients that minimize a $\chi^{2}$ expression of the type

\begin{equation}
\chi^{2} = \sum_{jk} \biggl[ T_{BEAST} - \sum_{i} (\alpha_i X_i)
\biggr]_j C^{-1}_{jk} \biggl[ T_{BEAST} - \sum_{i} (\alpha_i X_i)
\biggr]_k,
\end{equation}

\noindent where $C_{jk}$ is the covariance matrix of the BEAST
data. Since the noise pattern and the CMB anisotropy maps
are uncorrelated Gaussian variables with zero mean, uncorrelated
with the foreground templates, and also considering $n$ and
$T_{CMB}$ as noise, the $T_{BEAST}$ temperature
fluctuations correspond to a mapping of the fluctuations in the
distribution of the Galactic emission. $C_{jk}$ also accounts for
any chance alignment between CMB and the Galactic templates
which dominate the uncertainty in the coupling constants. Similar
analysis has been applied to data sets from different experiments
that characterize CMB fluctuations (Bennett et al. 1993; Kogut
et al. 1996a, b; de Oliveira-Costa et al. 1997, 1999; Hamilton
$\&$ Ganga 2001 and Mukherjee et al. 2002, 2003).

Considering the $X_i$ vectors as constants, the $C_{jk}$ matrix
is given by

\begin{equation}
%\begin{split}
C = \langle T_{BEAST}T_{BEAST}^T \rangle - \langle T_{BEAST} \rangle \langle
T_{BEAST}^T \rangle = \langle T_{CMB}T_{CMB}^T \rangle + \langle nn^T \rangle  = C_{CMB} + C_n, 
%\end{split}
\end{equation}

%$$
%C = \langle T_{BEAST}T_{BEAST}^T \rangle - \langle T_{BEAST} \rangle \langle
%T_{BEAST}^T \rangle
%  = \langle T_{CMB}T_{CMB}^T \rangle + \langle nn^T \rangle
%  = C_{CMB} + C_n,}
%$$ 

\vspace{0.3 cm}
\noindent the sum of the covariance matrix modeling the CMB signal
and the noise covariance matrix. Minimizing the $\chi^{2}$, the
best estimates of $\alpha_i$ are obtained as the solutions to
the system of equations

\begin{equation}
(X) C^{-1} (X)^T \tilde\alpha =
(X) C^{-1} (T_{BEAST})^T
\end{equation}

\noindent with a variance given by

\begin{equation}
\sigma_{\tilde\alpha}^2 = [(X) C^{-1} (X)^T]^{-1}.
\end{equation}

In Equations (4) and (5), the $X$ terms correspond to $(M \times N)$
arrays, where $M$ is the number of foreground templates
simultaneously analyzed and $N$ is the number of pixels in the
maps, and the $\tilde\alpha$ term corresponds to a $(M \times 1)$ vector
representing the unknown $\alpha_i$ parameters to be evaluated.

Finally, we estimated the level of Galactic contribution to the BEAST
maps by comparing the rms of the fluctuations in the Galactic templates,
scaled by their coupling constants ($\Delta T \equiv \tilde\alpha \cdot\sigma_{Gal}$),
to the rms of the BEAST temperature
fluctuations in the corresponding Ka- and Q-band maps. The rms values
follow directly from the estimates of the individual map variances

\begin{equation}
\sigma_{Gal, i}^2 = \frac{(X_i)^T (X_i)}{N},
\end{equation}

\noindent for each Galactic contaminant template and

\begin{equation}
\sigma_{BEAST}^2 = \frac{(T_{BEAST})^T (T_{BEAST})}{N},
\end{equation}

\noindent for the BEAST maps.

\section{ANALYSIS AND RESULTS}
The two BEAST maps at Ka- and Q-band were divided, along the RA
axis, in 24 sections of one-hour (RA), beginning at RA=0 h, each
with $\sim7900$ pixels, and were independently analyzed. To
characterize the correlation with the foreground templates, we
calculated the Pearson's linear correlation coefficient for the 24
sub-maps for each of the two BEAST maps. The results are
presented in Tables 1 to 6 and Figures 3 to 5. We should point out
that the templates we used do not necessarily trace the behavior
of the individual Galactic foreground contaminants at BEAST
frequencies, even though the generally low values of the
correlation coefficients over the 1-hour RA section may suggest
this.

For the free-free emission, in at least two regions in both
BEAST maps, identified as regions 3 (3\ts h $\leq$ RA $<4$\ts h) and
4 (4\ts h $\leq$ RA $<5$\ts h) in Tables 1 and 2, the value of 
the correlation coefficient was
above 0.45 for both bands (p\ts $<$\ts 0.0001), high enough to
consider them as highly contaminated by the free-free emission. It
is possible to see, in the same tables, that the percentage of
temperature contribution of the free-free emission (as traced by
the H$\alpha$ template) to the corresponding BEAST map regions
varies between 49.0$\%$ and 56.5$\%$ (Ka-band) and between
48.1$\%$ and 64.4$\%$ (Q-band).

In the same tables, we present the values of the $\alpha$
parameter as obtained from Equation 4. For the regions mentioned
above, the $\alpha$ parameter varies from ($6.4\pm0.4$) $\mu$K/R
to ($8.3\pm0.4$) $\mu$K/R for the Ka-band, and from ($3.3\pm0.2$)
$\mu$K/R to ($4.1\pm0.2$) $\mu$K/R for the Q-band. These values agree
with the theoretical results presented by Dickinson et al. (2003),
for an electronic temperature T$_e \sim 10^4$ K.

For synchrotron emission, the correlation coefficient was always 
below 0.15, except for region 16 (16\ts h $\leq$ RA $<17$\ts h) 
in the Ka-band and regions 16 (16\ts h $\leq$ RA $<17$\ts h) and 23
(23\ts h $\leq$ RA $<24$\ts h) in Q-band, in Tables 3 and 4. 
For these regions the constribution in temperature corresponds to
16.8 $\%$ in the Ka-band, and varies between 16.7$\%$ and 23.0$\%$ 
in the Q-band.

For thermal dust emission, the same two regions as in the free-free
emission case in both BEAST maps, identified as regions 3 and 4 in
Tables 5 and 6, present values of the correlation coefficient
above 0.42 for both bands (p\ts $<$\ts 0.0001), high enough to
consider them as highly contaminated by the thermal dust emission.
The percentage of temperature contribution of the thermal dust
emission to the corresponding BEAST map regions varies between
42.1$\%$ and 67.4$\%$ (Ka-band), and between 47.0$\%$ and 67.5$\%$
(Q-band).

The $\alpha$ parameter values obtained for the
regions mentioned above vary from ($30.1\pm2.1$) $\mu$K/(MJy/sr)
to ($45.0.0\pm2.0$) $\mu$K/(MJy/sr) for the Ka-band and from
($14.6\pm1.2$) $\mu$K/(MJy/sr) to ($24.0\pm1.1$) $\mu$K/(MJy/sr)
for the Q-band.

Due to the large number of pixels in the BEAST maps, we are able
to determine the $\alpha$ values corresponding to different
regions of the Galaxy, which allow us to take into account spatial
variations in the Galaxy foreground emission. For instance, we focused on
one interesting feature found in BEAST maps, which we called ``the
bar" (Meinhold et al. 2003), centered around RA = 4.04 h ($60.6
^\circ$) and $\delta$ = $36.2^\circ$, shown in Figure 6, and
performed the same calculation for the region located between 58.3$^\circ<$ RA
$<62.7^\circ$ and 34.7$^\circ< \delta <38.0^\circ$ obtaining as
results ($7.51\pm0.31$) $\mu$K/R at 30 GHz and ($3.76\pm0.17$)
$\mu$K/R at 41.5 GHz. The corresponding Pearson's
correlation coefficients were 0.648 (Ka-band) and 0.623 (Q-band)
(875 pixels; p\ts $<$\ts 0.0001).

In order to determine regions of the sky to be avoided for BEAST
CMB analysis due to Galactic foreground contamination, we applied
the method outlined in the previous section to evaluate the
Galactic foreground contribution in our data set. BEAST beams
crossed twice the Galactic Plane in the observational campaign at
WMRS. This allows us to examine in detail the extent of the
Galactic contamination in our data. We began with a 2.5$^\circ$
cut above and below the Galactic Plane and estimated the
individual contribution of each foreground component in the
remaining part of the BEAST maps. We set a $\mid$b$\mid = 5^\circ$
step for further cuts. Tables 7 and 8 show the results of this
procedure. The number of pixels left in the BEAST maps is
presented in the second column of Tables 7 and 8. Columns 4, 6,
and 8, in both tables, present the percentual temperature
contribution of each foreground component, estimated from the
templates, while columns 3, 5, and 7 show the corresponding $\alpha$
parameters for the three foregrounds after each sucessive Galactic
Plane remotion.
It is clear from these tables that the Galactic
foreground contamination is negligible after a $\mid$b$\mid =
17.5^\circ$ Galactic cut. This cut was then applied to our data
set in order to generate BEAST CMB maps (Meinhold et al. 2003)
and estimate CMB power spectrum (O'Dwyer et al. 2004).

\acknowledgments We want to thank the White Mountain Research
Station staff for the very important support during BEAST
operation. 
This work was funded by NASA grants NAG5-4078, NAG5-9073,
and NAG5-4185, and by NSF grants 9813920 and 0118297. 
In addition we were supported by the White Mountain Research Station, 
the California Space Institute (CalSpace) and the UCSB Office of Research.
The development and operations of BEAST were supported
by NASA Office of Space Sciences, the National Science Foundation,
University of California White Mountain Research Station, and the
California Space Institute (CalSpace). 
J.M. was supported by FAPESP grants 01/13235-9 and 02/04871-1. 
T.V. was partially supported by FAPESP grant 00/06770-2 and by CNPq grants
466184/00-0 and 302266/88-7-FA. 
C.A.W. was partially supported by CNPq grant 300409/97-4-FA and FAPESP 
grants 00/06770-2 and 96/06501-4. 
N.F. and A.P. were partially supported by CNPq grant 470531/2001-0. 
Some of the results in this paper have been derived using the HEALPix 
(G\'{o}rski et al. 1999) package.

\clearpage

%Figure Ka-band BEAST map ---------------------------------------
\begin{figure}
\plotone{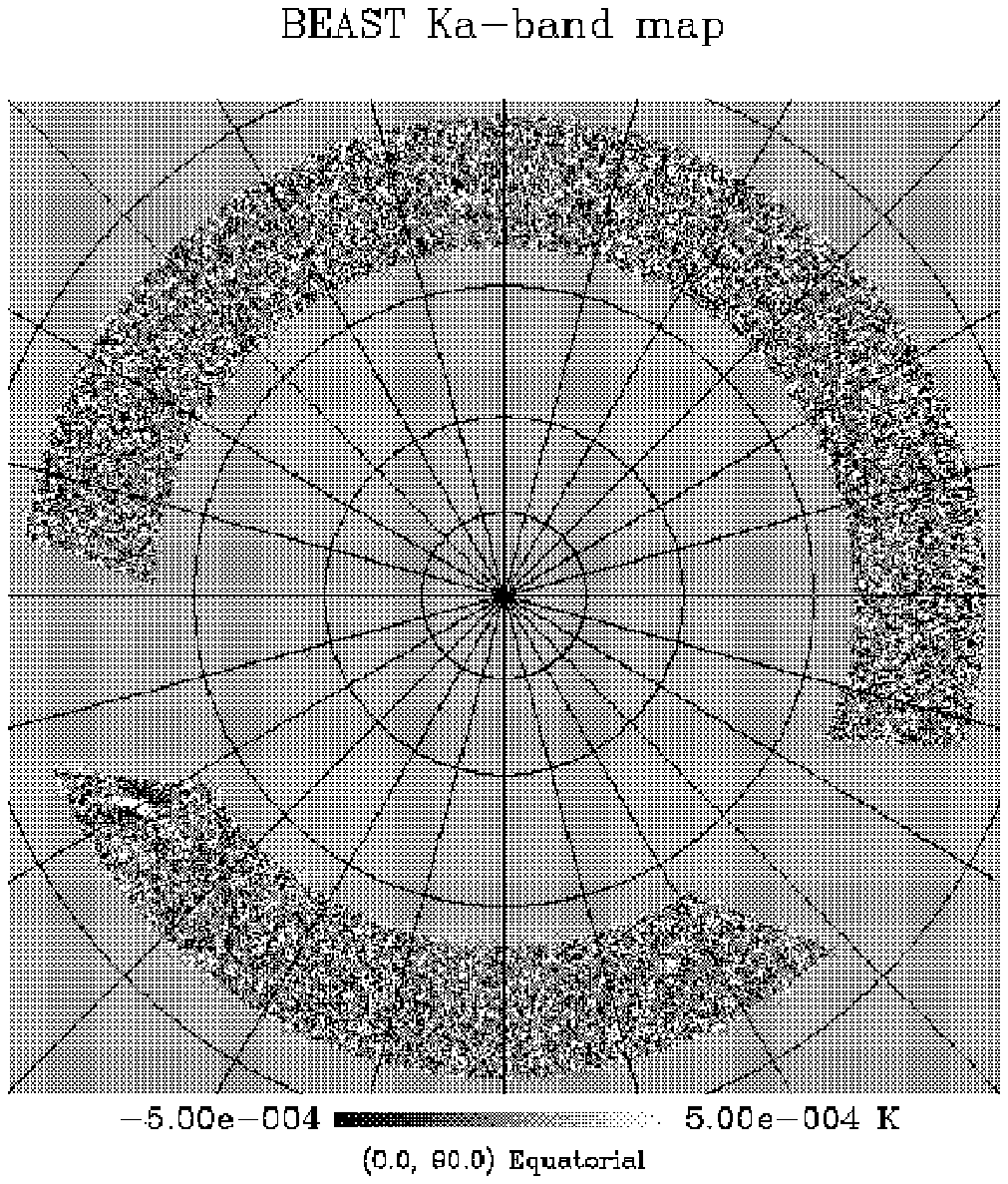}
\caption{BEAST Ka-band map in gnomonic projection. RA\ts =\ts 0\ts
h is at the bottom, increasing clockwise, and the North Celestial
Pole is at the center. Each graticule division corresponds to
15$^\circ$ x 15$^\circ$ (RA, $\delta$).}
\end{figure}

%Figure Q-band BEAST map ---------------------------------------
\begin{figure}
\plotone{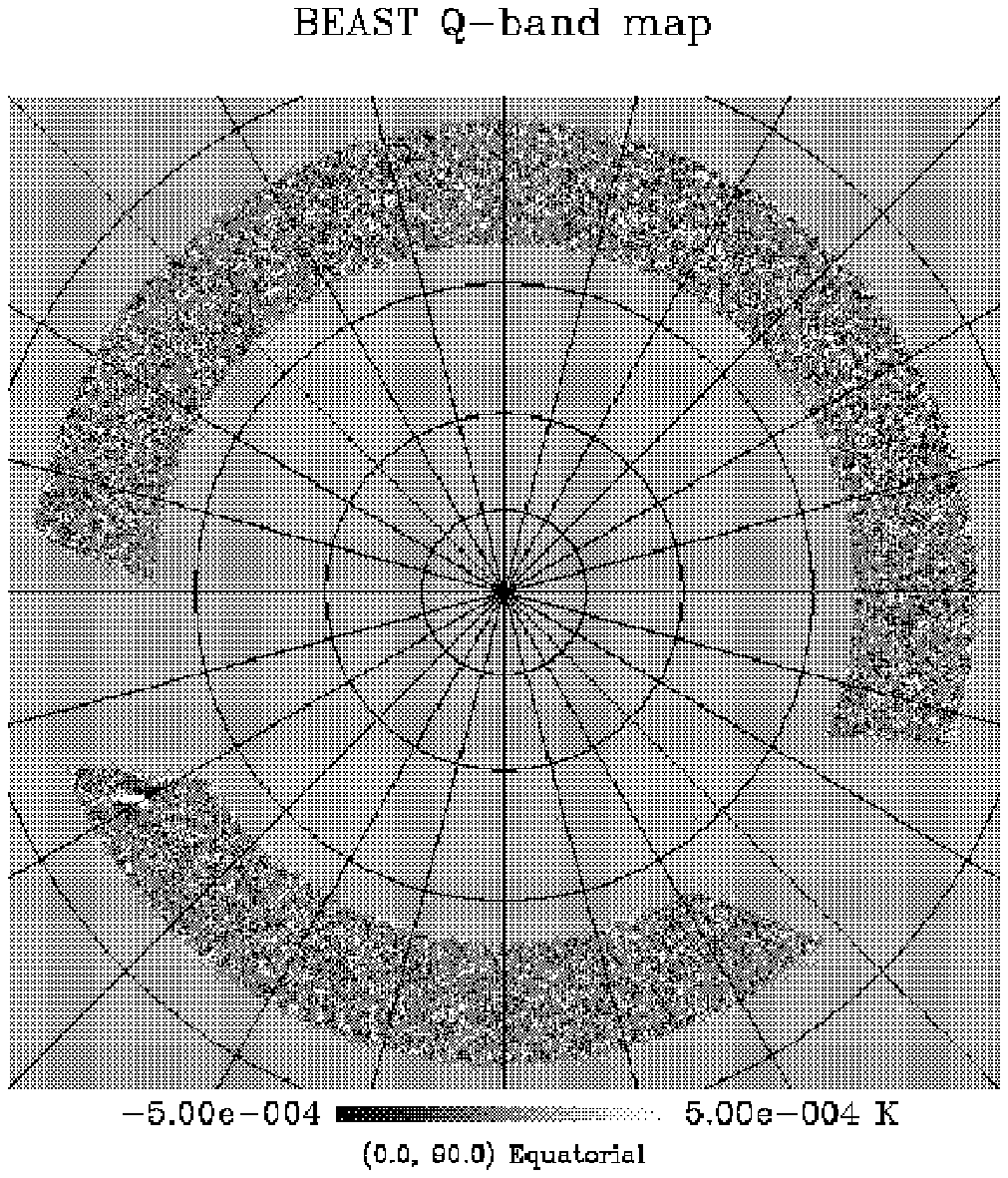}
\caption{BEAST Q-band map in gnomonic projection. RA\ts =\ts 0\ts
h is at the bottom, increasing clockwise, and the North Celestial
Pole is at the center. Each graticule division corresponds to
15$^\circ$ x 15$^\circ$ (RA, $\delta$).}
\end{figure}

\clearpage

%Figure Pearson Ka- and Q-Halpha ---------------------------------------
\begin{figure}
\plotone{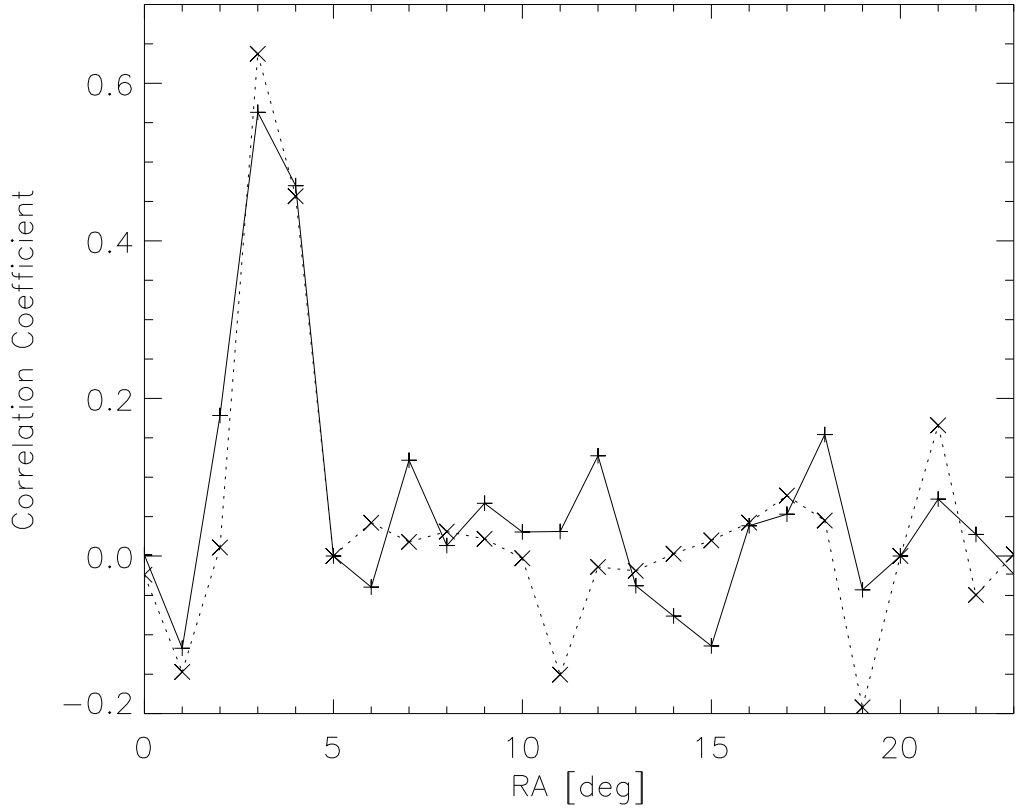}
\caption{Pearson's coefficients for the linear correlation between
BEAST Ka-band map and H$\alpha$ template (dotted line), and BEAST
Q-band map and H$\alpha$ template (solid line).}
\end{figure}

%Figure Pearson Ka- and Q-Sync ---------------------------------------
\begin{figure}
\plotone{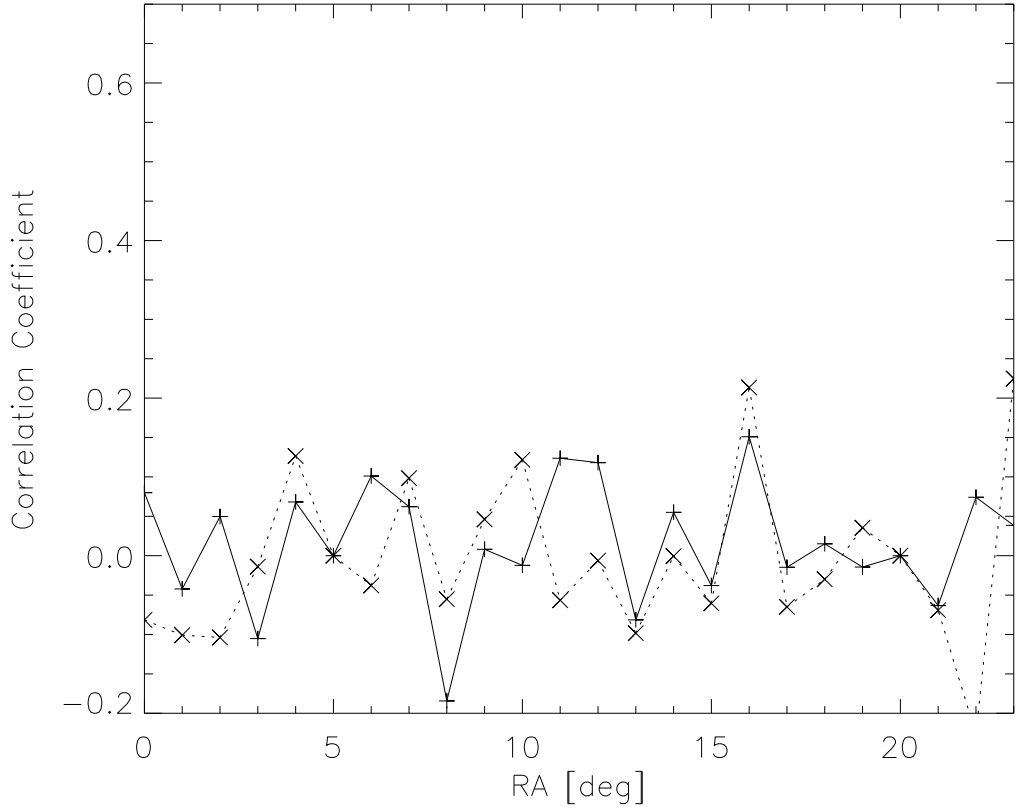}
\caption{Pearson's coefficients for the linear correlation between
BEAST Ka-band map and synchrotron 408 MHz template (dotted line),
and BEAST Q-band map and synchrotron 408 MHz template (solid
line).}
\end{figure}

%Figure Pearson Ka- and Q-Dust ---------------------------------------
\begin{figure}
\plotone{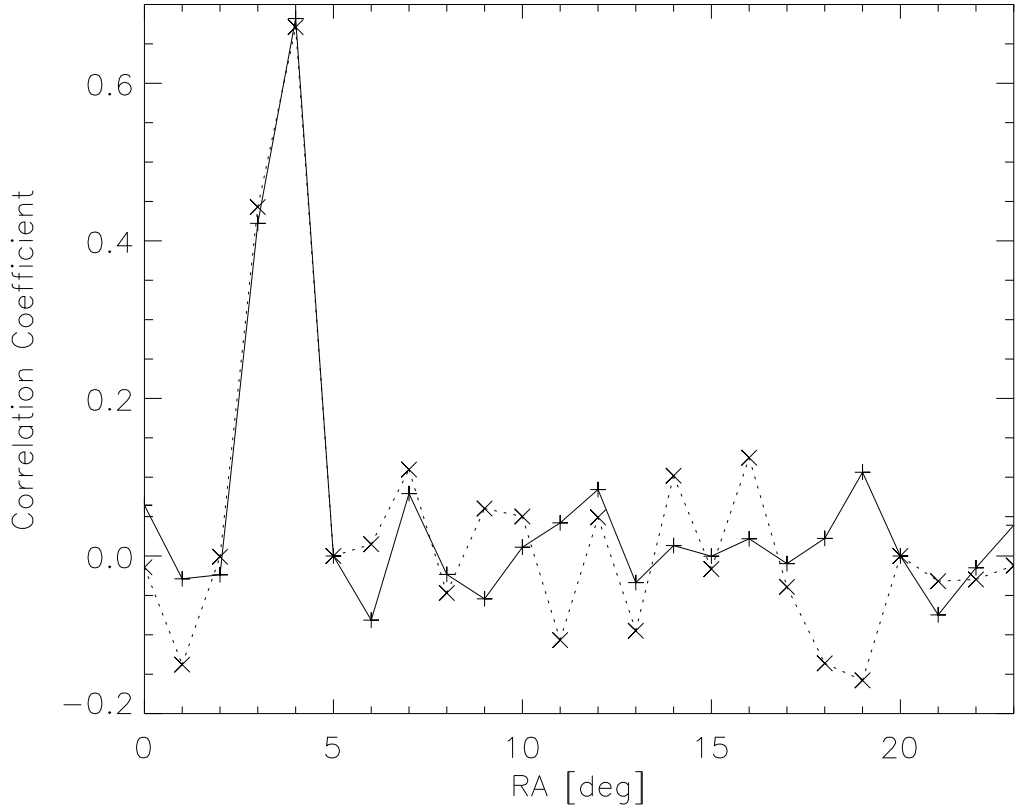}
\caption{Pearson's coefficients for the linear correlation between
BEAST Ka-band map and dust template (dotted line), and BEAST Q-band
map and dust template (solid line).}
\end{figure}

\clearpage

%Figure The BAR ---------------------------------------
\begin{figure}
\plottwo{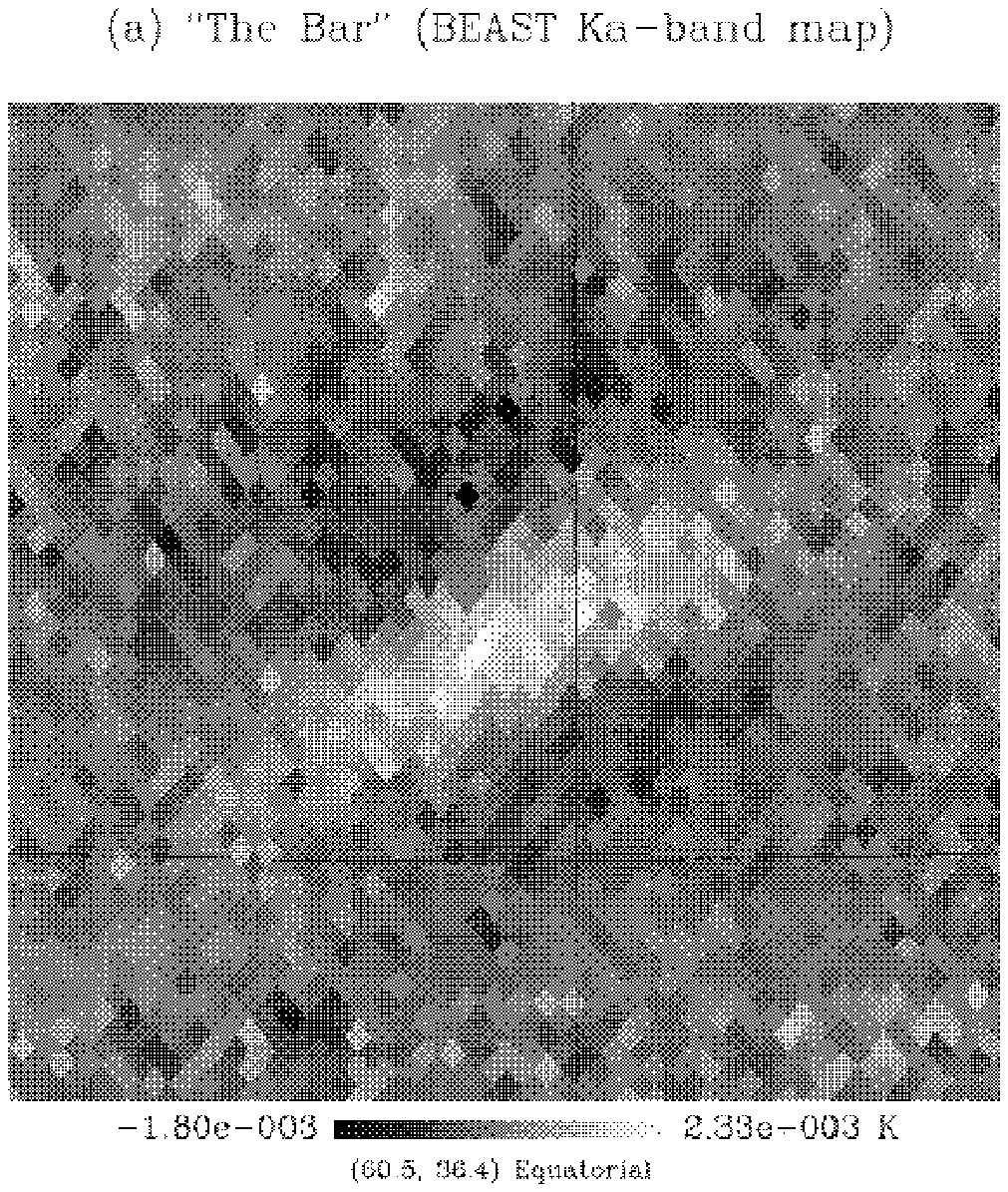}{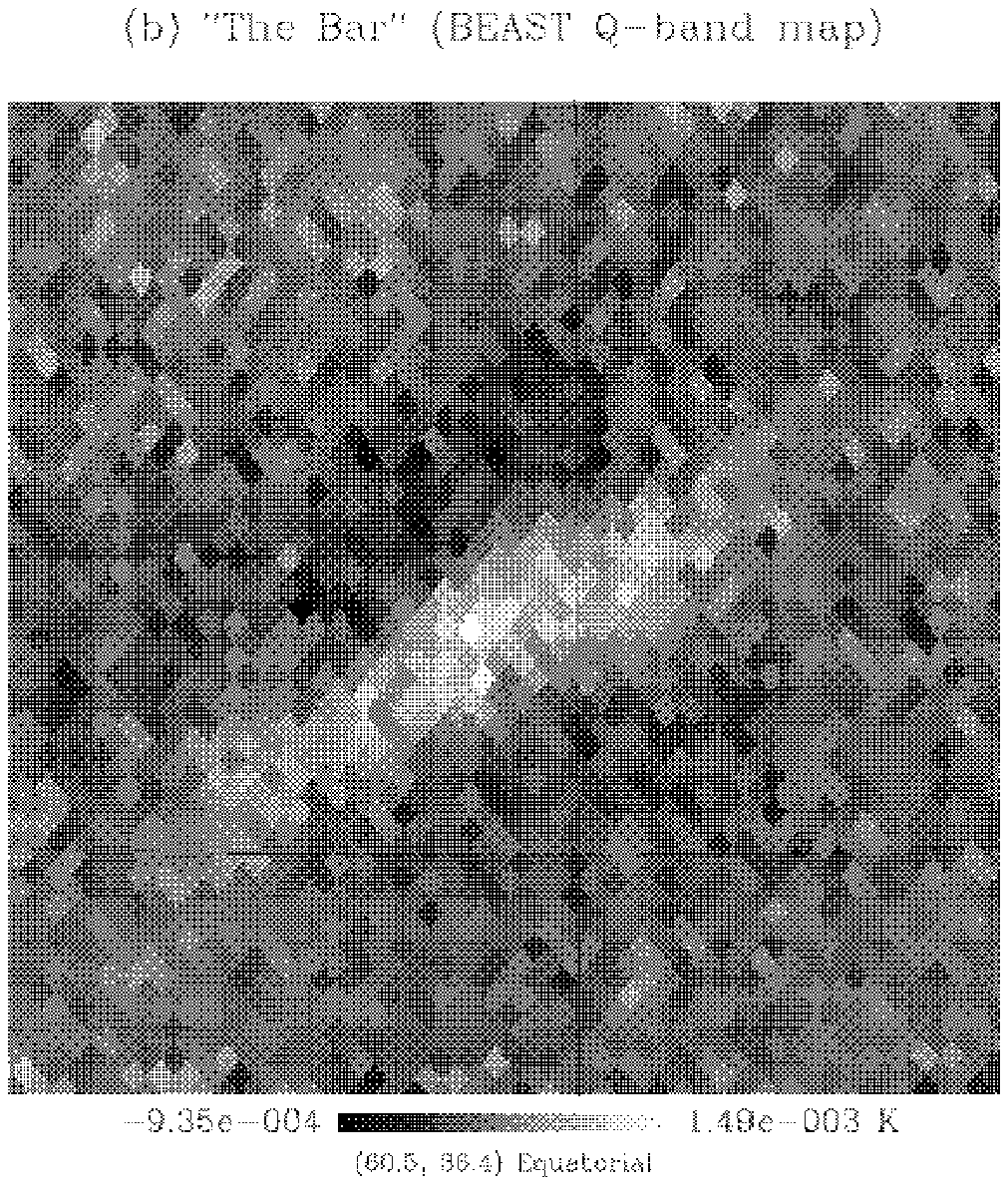}
\epsscale{2.22}
\plottwo{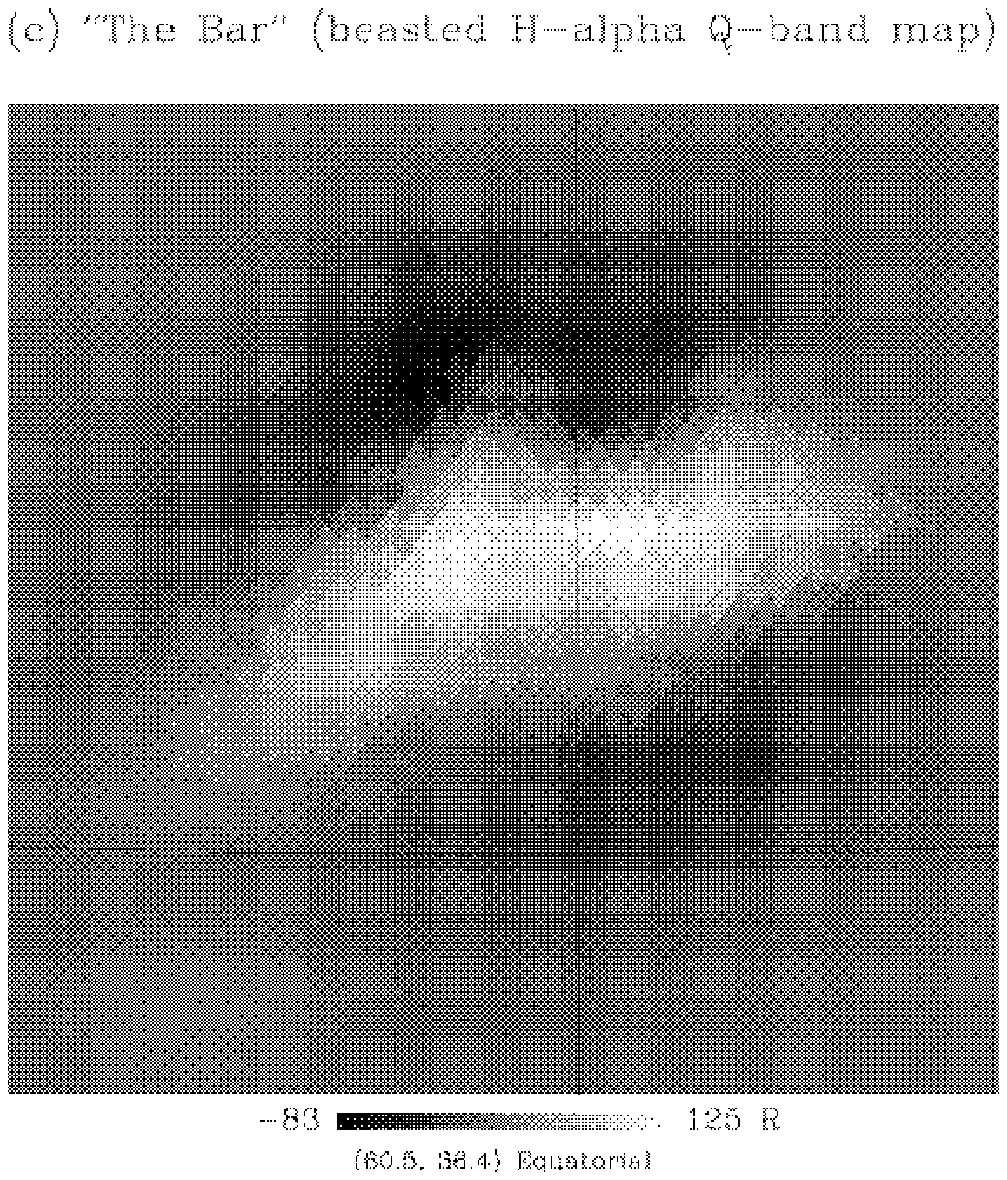}{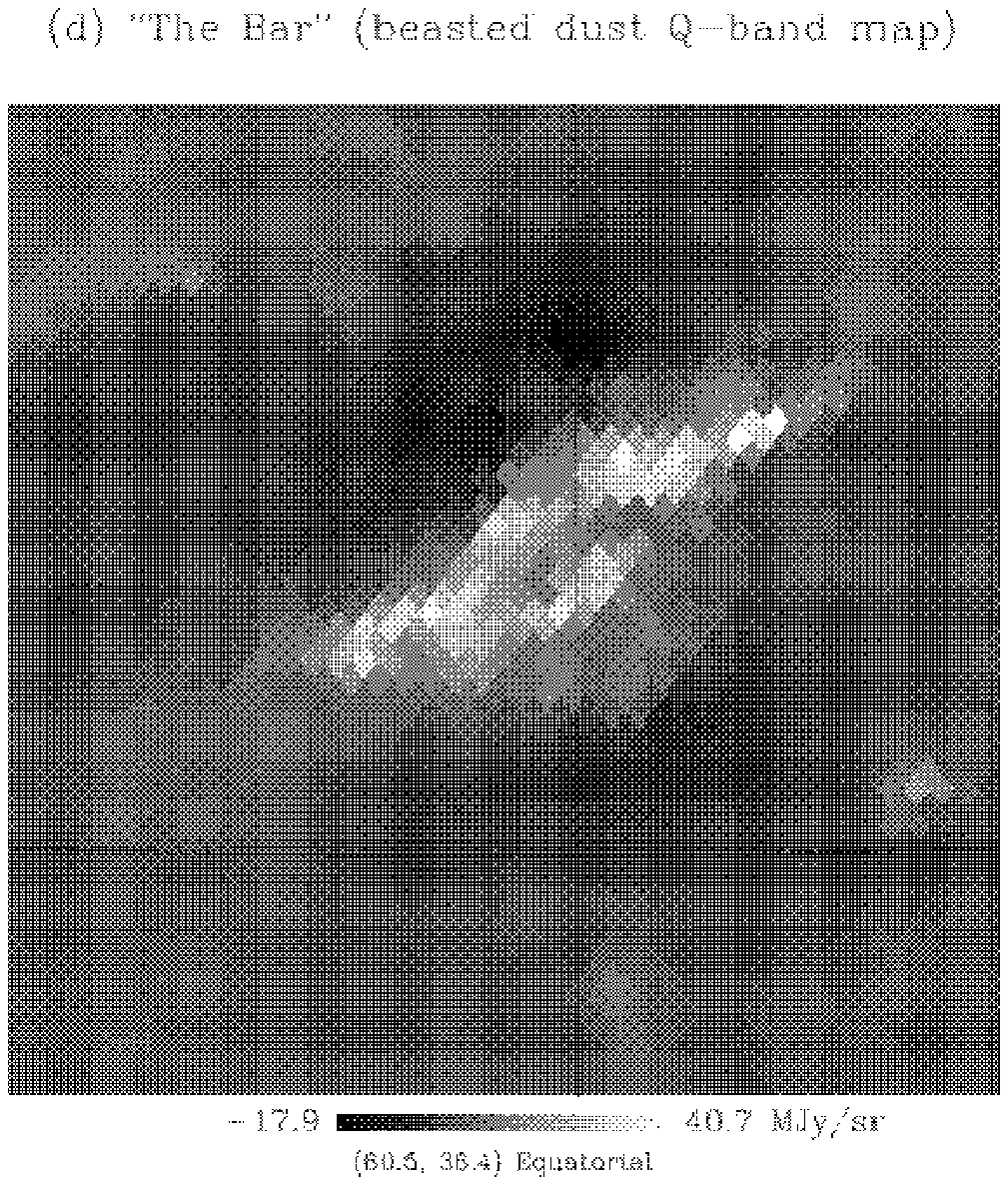}
\caption{``The bar" as seen in the (a) BEAST Ka- and (b) BEAST
Q-band maps and in the (c) ``beasted" H$\alpha$ and (d) ``beasted"
dust templates. The images are centered on RA\ts=\ts 60.5$^\circ$
and $\delta$\ts=\ts 36.4$^\circ$.}
\end{figure}

\clearpage

%Table: Ka_BEAST and Halpha coefficient -----------------------------------------
\begin{deluxetable}{ccccccccc}
\tabletypesize{\scriptsize}
\tablecaption{Correlation results between BEAST Ka-band map and H$\alpha$ template}   %. \label{tbl-1}}
\tablewidth{0pt}
\tablehead{
\colhead{Section} & 
\colhead{Number of} & 
\colhead{Correlation}   &
\colhead{Probability} &
\colhead{$\tilde\alpha$}  & 
\colhead{$\sigma_{\tilde\alpha}$} & 
\colhead{$\sigma_{Gal}$} &
\colhead{$\Delta T \equiv \tilde\alpha \cdot\sigma_{Gal}$}     & 
\colhead{$\Delta T / \sigma_B$} \\
\colhead{} &
\colhead{pixels} &
\colhead{Coeff.} &
\colhead{[$\%$]} &
\colhead{[$\mu${K/R}]}  & 
\colhead{[$\mu${K/R}]} & 
\colhead{[$R$]} & 
\colhead{[$\mu$K]} & 
\colhead{[$\%$]}
}
\startdata
  0  &     7951  &   0.0013 &    45.41  &      -0.334 &       9.848 &       0.343  &    -0.11  &    -0.14 \\
  1  &     7952  &  -0.1171 & $<$ 0.01  &     -50.643 &      18.460 &       0.214  &   -10.84  &   -12.00 \\
  2  &     7951  &   0.1784 & $<$ 0.01  &      50.981 &      12.498 &       0.310  &    15.78  &    17.08 \\
  3  &     7780  &   0.5631 & $<$ 0.01  &       8.318 &       0.440 &       9.241  &    76.87  &    56.48 \\
  4  &     2009  &   0.4701 & $<$ 0.01  &       6.393 &       0.391 &      20.624  &   131.84  &    49.00 \\
  6  &     5467  &  -0.0395 &     0.17  &     -25.822 &      28.888 &       0.181  &    -4.68  &    -4.95 \\
  7  &     7952  &   0.1215 & $<$ 0.01  &     101.153 &      41.827 &       0.102  &    10.34  &    12.00 \\
  8  &     7951  &   0.0133 &    11.80  &      19.786 &      71.587 &       0.058  &     1.14  &     1.18 \\
  9  &     7905  &   0.0669 & $<$ 0.01  &     302.061 &     127.176 &       0.032  &     9.55  &     9.57 \\
 10  &     7952  &   0.0305 &     0.33  &     112.272 &     106.123 &       0.038  &     4.32  &     4.34 \\
 11  &     7951  &   0.0311 &     0.28  &     122.695 &     133.724 &       0.031  &     3.83  &     3.68 \\
 12  &     7905  &   0.1272 & $<$ 0.01  &     529.030 &     159.984 &       0.025  &    13.13  &    13.00 \\
 13  &     7951  &  -0.0377 &     0.04  &    -110.799 &     101.603 &       0.040  &    -4.46  &    -3.53 \\
 14  &     7949  &  -0.0763 & $<$ 0.01  &    -195.850 &     112.354 &       0.038  &    -7.43  &    -6.58 \\
 15  &     7899  &  -0.1142 & $<$ 0.01  &    -303.971 &     112.344 &       0.038  &   -11.46  &   -11.16 \\
 16  &     7947  &   0.0383 &     0.03  &     113.628 &      80.361 &       0.056  &     6.36  &     5.34 \\
 17  &     7944  &   0.0531 & $<$ 0.01  &      46.845 &      44.450 &       0.098  &     4.60  &     4.43 \\
 18  &     7821  &   0.1542 & $<$ 0.01  &      54.151 &      13.304 &       0.316  &    17.09  &    14.58 \\
 19  &     3211  &  -0.0430 &     0.75  &      -4.796 &      13.634 &       0.493  &    -2.36  &    -2.09 \\
 21  &     4134  &   0.0722 & $<$ 0.01  &       5.826 &       3.541 &       1.449  &     8.44  &     8.74 \\
 22  &     7952  &   0.0274 &     0.73  &       1.547 &       1.219 &       2.883  &     4.46  &     4.98 \\
 23  &     7905  &  -0.0229 &     2.08  &     -17.183 &      16.594 &       0.219  &    -3.76  &    -4.41
\enddata
\end{deluxetable}

\clearpage

%Table: Q_BEAST and Halpha coefficient -----------------------------------------
\begin{deluxetable}{ccccccccc}
\tabletypesize{\scriptsize}
\tablecaption{Correlation results between BEAST Q-band map and H$\alpha$ template}%. \label{tbl-1}}
\tablewidth{0pt}
\tablehead{
\colhead{Section} & 
\colhead{Number of} & 
\colhead{Correlation}   &
\colhead{Probability} &
\colhead{$\tilde\alpha$}  & 
\colhead{$\sigma_{\tilde\alpha}$} & 
\colhead{$\sigma_{Gal}$} &
\colhead{$\Delta T \equiv \tilde\alpha \cdot\sigma_{Gal}$} & 
\colhead{$\Delta T / \sigma_B$} \\
\colhead{} &
\colhead{pixels} &
\colhead{Coeff.} &
\colhead{[$\%$]} &
\colhead{[$\mu${K/R}]}  & 
\colhead{[$\mu${K/R}]} & 
\colhead{[$R$]} & 
\colhead{[$\mu$K]} & 
\colhead{[$\%$]}
}
\startdata
  0  &     7866  &  -0.0237 &     1.77  &      -3.628 &       7.307 &       0.231  &    -0.84  &    -1.84 \\
  1  &     7866  &  -0.1471 & $<$ 0.01  &     -29.253 &      10.543 &       0.223  &    -6.53  &   -13.56 \\
  2  &     7866  &   0.0108 &    16.87  &      -0.512 &       6.510 &       0.314  &    -0.16  &    -0.34 \\
  3  &     7723  &   0.6373 & $<$ 0.01  &       4.093 &       0.239 &       9.907  &    40.54  &    64.39 \\
  4  &     2059  &   0.4567 & $<$ 0.01  &       3.284 &       0.211 &      21.961  &    72.13  &    48.08 \\
  6  &     5361  &   0.0422 &     0.10  &      13.347 &      14.109 &       0.193  &     2.58  &     5.52 \\
  7  &     7866  &   0.0179 &     5.65  &      31.502 &      21.608 &       0.108  &     3.40  &     6.97 \\
  8  &     7866  &   0.0311 &     0.29  &      50.767 &      41.305 &       0.061  &     3.07  &     6.55 \\
  9  &     7820  &   0.0219 &     2.67  &     122.951 &      72.972 &       0.033  &     4.07  &     7.90 \\
 10  &     7866  &  -0.0029 &    39.81  &      21.357 &      57.900 &       0.040  &     0.86  &     1.49 \\
 11  &     7866  &  -0.1505 & $<$ 0.01  &    -215.064 &      68.664 &       0.032  &    -6.92  &   -15.75 \\
 12  &     7820  &  -0.0139 &    11.02  &     -92.942 &      92.754 &       0.025  &    -2.34  &    -3.74 \\
 13  &     7866  &  -0.0189 &     4.72  &     -56.237 &      59.903 &       0.042  &    -2.35  &    -4.09 \\
 14  &     7866  &   0.0028 &    40.15  &     -48.228 &      64.271 &       0.039  &    -1.89  &    -2.90 \\
 15  &     7820  &   0.0198 &     3.99  &     -26.230 &      63.992 &       0.039  &    -1.03  &    -1.65 \\
 16  &     7866  &   0.0424 &     0.01  &      11.437 &      42.818 &       0.060  &     0.69  &     0.89 \\
 17  &     7866  &   0.0767 & $<$ 0.01  &      52.133 &      24.553 &       0.103  &     5.36  &     7.74 \\
 18  &     7820  &   0.0450 & $<$ 0.01  &       1.002 &       7.716 &       0.332  &     0.33  &     0.64 \\
 19  &     3135  &  -0.1921 & $<$ 0.01  &     -26.138 &       7.276 &       0.518  &   -13.55  &   -24.03 \\
 21  &     4187  &   0.1659 & $<$ 0.01  &       5.066 &       1.926 &       1.510  &     7.65  &    17.38 \\
 22  &     7866  &  -0.0494 & $<$ 0.01  &      -0.202 &       0.727 &       2.953  &    -0.60  &    -1.26 \\
 23  &     7820  &   0.0020 &    42.92  &       5.084 &       9.601 &       0.225  &     1.15  &     2.38
\enddata
\end{deluxetable}

\clearpage

%Table: Ka_BEAST and Synchrotron coefficient -----------------------------------------
\begin{deluxetable}{ccccccccc}
\tabletypesize{\scriptsize}
\tablecaption{Correlation results between BEAST Ka-band map and synchrotron template}%. \label{tbl-1}}
\tablewidth{0pt}
\tablehead{
\colhead{Section} & 
\colhead{Number of} & 
\colhead{Correlation}   &
\colhead{Probability} &
\colhead{$\tilde\alpha$}  & 
\colhead{$\sigma_{\tilde\alpha}$} & 
\colhead{$\sigma_{Gal}$} &
\colhead{$\Delta T \equiv \tilde\alpha \cdot\sigma_{Gal}$}     & 
\colhead{$\Delta T / \sigma_B$} \\
\colhead{} &
\colhead{pixels} &
\colhead{Coeff.} &
\colhead{[$\%$]} &
\colhead{[$\mu${K/K}]}  & 
\colhead{[$\mu${K/K}]} & 
\colhead{[$K$]} & 
\colhead{[$\mu$K]} & 
\colhead{[$\%$]}
}
\startdata
  0  &     7951  &   0.0800 & $<$ 0.01  &      34.971 &      19.561 &       0.190  &     6.64  &     8.25 \\
  1  &     7952  &  -0.0422 &     0.01  &     -21.833 &      27.558 &       0.133  &    -2.91  &    -3.23 \\
  2  &     7951  &   0.0497 & $<$ 0.01  &      26.748 &      22.201 &       0.172  &     4.60  &     4.98 \\
  3  &     7780  &  -0.1052 & $<$ 0.01  &     -88.337 &      23.274 &       0.166  &   -14.65  &   -10.76 \\
  4  &     2009  &   0.0683 &     0.11  &     124.116 &      36.736 &       0.204  &    25.35  &     9.42 \\
  6  &     5467  &   0.1011 & $<$ 0.01  &      56.083 &      28.984 &       0.175  &     9.83  &    10.40 \\
  7  &     7952  &   0.0622 & $<$ 0.01  &      27.298 &      27.256 &       0.154  &     4.21  &     4.88 \\
  8  &     7951  &  -0.1843 & $<$ 0.01  &    -101.841 &      25.641 &       0.161  &   -16.39  &   -16.87 \\
  9  &     7905  &   0.0081 &    23.59  &       0.660 &      23.600 &       0.169  &     0.11  &     0.11 \\
 10  &     7952  &  -0.0123 &    13.66  &      -5.424 &      28.091 &       0.149  &    -0.81  &    -0.81 \\
 11  &     7951  &   0.1237 & $<$ 0.01  &      65.980 &      28.340 &       0.148  &     9.73  &     9.35 \\
 12  &     7905  &   0.1181 & $<$ 0.01  &      88.272 &      27.529 &       0.144  &    12.71  &    12.57 \\
 13  &     7951  &  -0.0814 & $<$ 0.01  &     -52.196 &      25.107 &       0.162  &    -8.46  &    -6.69 \\
 14  &     7949  &   0.0550 & $<$ 0.01  &      31.858 &      26.020 &       0.163  &     5.20  &     4.60 \\
 15  &     7899  &  -0.0379 &     0.04  &     -25.449 &      25.558 &       0.159  &    -4.06  &    -3.95 \\
 16  &     7947  &   0.1510 & $<$ 0.01  &     126.873 &      27.972 &       0.157  &    19.98  &    16.78 \\
 17  &     7944  &  -0.0147 &     9.50  &       3.388 &      23.043 &       0.189  &     0.64  &     0.62 \\
 18  &     7821  &   0.0151 &     9.07  &       4.883 &      15.154 &       0.280  &     1.37  &     1.17 \\
 19  &     3211  &  -0.0144 &    20.66  &       1.086 &       4.729 &       1.382  &     1.50  &     1.33 \\
 21  &     4134  &  -0.0633 & $<$ 0.01  &     -26.173 &      18.433 &       0.284  &    -7.44  &    -7.70 \\
 22  &     7952  &   0.0742 & $<$ 0.01  &      22.775 &      15.579 &       0.227  &     5.17  &     5.77 \\
 23  &     7905  &   0.0385 &     0.03  &      12.597 &      20.843 &       0.183  &     2.30  &     2.70 
\enddata
\end{deluxetable}

\clearpage

%Table: Q_BEAST and Synchrotron coefficient -----------------------------------------
\begin{deluxetable}{ccccccccc}
\tabletypesize{\scriptsize}
\tablecaption{Correlation results between BEAST Q-band map and
synchrotron template}%. \label{tbl-1}}
\tablewidth{0pt}
\tablehead{
\colhead{Section} & 
\colhead{Number of} & 
\colhead{Correlation}   &
\colhead{Probability} &
\colhead{$\tilde\alpha$}  & 
\colhead{$\sigma_{\tilde\alpha}$} & 
\colhead{$\sigma_{Gal}$} &
\colhead{$\Delta T \equiv \tilde\alpha \cdot\sigma_{Gal}$}     & 
\colhead{$\Delta T / \sigma_B$} \\
\colhead{} &
\colhead{pixels} &
\colhead{Coeff.} &
\colhead{[$\%$]} &
\colhead{[$\mu${K/K}]}  & 
\colhead{[$\mu${K/K}]} & 
\colhead{[$K$]} & 
\colhead{[$\mu K$]} & 
\colhead{[$\%$]}
}
\startdata
  0  &     7866  &  -0.0816 & $<$ 0.01  &     -22.379 &      11.054 &       0.199  &    -4.45  &    -9.79 \\
  1  &     7866  &  -0.1009 & $<$ 0.01  &     -39.178 &      15.506 &       0.135  &    -5.30  &   -11.00 \\
  2  &     7866  &  -0.1038 & $<$ 0.01  &     -28.106 &      12.827 &       0.177  &    -4.98  &   -10.40 \\
  3  &     7723  &  -0.0139 &    11.12  &       2.197 &      11.932 &       0.171  &     0.37  &     0.60 \\
  4  &     2059  &   0.1265 & $<$ 0.01  &     117.535 &      19.753 &       0.219  &    25.79  &    17.20 \\
  6  &     5361  &  -0.0378 &     0.28  &      -9.900 &      16.454 &       0.180  &    -1.78  &    -3.82 \\
  7  &     7866  &   0.0985 & $<$ 0.01  &      43.436 &      14.335 &       0.160  &     6.97  &    14.29 \\
  8  &     7866  &  -0.0551 & $<$ 0.01  &     -13.540 &      14.155 &       0.164  &    -2.23  &    -4.74 \\
  9  &     7820  &   0.0461 & $<$ 0.01  &      19.541 &      12.704 &       0.174  &     3.40  &     6.61 \\
 10  &     7866  &   0.1218 & $<$ 0.01  &      47.519 &      15.324 &       0.156  &     7.44  &    12.91 \\
 11  &     7866  &  -0.0565 & $<$ 0.01  &     -26.170 &      14.895 &       0.156  &    -4.08  &    -9.29 \\
 12  &     7820  &  -0.0063 &    28.90  &       1.703 &      16.089 &       0.143  &     0.24  &     0.39 \\
 13  &     7866  &  -0.0983 & $<$ 0.01  &     -38.581 &      14.201 &       0.166  &    -6.39  &   -11.13 \\
 14  &     7866  &  -0.0006 &    47.78  &       9.103 &      14.197 &       0.170  &     1.55  &     2.38 \\
 15  &     7820  &  -0.0605 & $<$ 0.01  &     -24.101 &      14.103 &       0.167  &    -4.02  &    -6.41 \\
 16  &     7866  &   0.2135 & $<$ 0.01  &      79.253 &      16.046 &       0.163  &    12.89  &    16.70 \\
 17  &     7866  &  -0.0650 & $<$ 0.01  &      -5.030 &      12.205 &       0.200  &    -1.00  &    -1.45 \\
 18  &     7820  &  -0.0299 &     0.41  &      -8.900 &       8.736 &       0.297  &    -2.64  &    -5.09 \\
 19  &     3135  &   0.0355 &     2.35  &       0.257 &       3.491 &       1.069  &     0.27  &     0.49 \\
 21  &     4187  &  -0.0692 & $<$ 0.01  &     -13.168 &      10.879 &       0.296  &    -3.90  &    -8.86 \\
 22  &     7866  &  -0.2265 & $<$ 0.01  &     -49.490 &       9.234 &       0.228  &   -11.26  &   -23.78 \\
 23  &     7820  &   0.2248 & $<$ 0.01  &      58.077 &      11.559 &       0.191  &    11.08  &    23.05
\enddata
\end{deluxetable}

\clearpage

%Table: Ka_BEAST and Dust coefficient -----------------------------------------
\begin{deluxetable}{ccccccccc}
\tabletypesize{\scriptsize}
\tablecaption{Correlation results between BEAST Ka-band map and
dust template}%. \label{tbl-1}}
\tablewidth{0pt}
\tablehead{
\colhead{Section} & 
\colhead{Number of} & 
\colhead{Correlation}   &
\colhead{Probability} &
\colhead{$\tilde\alpha$}  & 
\colhead{$\sigma_{\tilde\alpha}$} & 
\colhead{$\sigma_{Gal}$} &
\colhead{$\Delta T \equiv \tilde\alpha \cdot\sigma_{Gal}$}     & 
\colhead{$\Delta T / \sigma_B$} \\
\colhead{} &
\colhead{pixels} &
\colhead{Coeff.} &
\colhead{[$\%$]} &
\colhead{[$\mu${K/(MJy/sr)}]}  & 
\colhead{[$\mu${K/(MJy/sr)}]} & 
\colhead{[$MJy/sr$]} & 
\colhead{[$\mu$K]} & 
\colhead{[$\%$]}
}
\startdata
  0  &     7951  &   0.0644 & $<$ 0.01  &      14.201 &      10.097 &       0.351  &     4.98  &     6.19 \\
  1  &     7952  &  -0.0289 &     0.50  &     -10.193 &      23.609 &       0.164  &    -1.67  &    -1.85 \\
  2  &     7951  &  -0.0238 &     1.69  &      -6.329 &      18.616 &       0.211  &    -1.33  &    -1.44 \\
  3  &     7780  &   0.4222 & $<$ 0.01  &      30.055 &       2.126 &       1.905  &    57.26  &    42.07 \\
  4  &     2009  &   0.6822 & $<$ 0.01  &      45.005 &       1.952 &       4.029  &   181.31  &    67.38 \\
  6  &     5467  &  -0.0814 & $<$ 0.01  &     -17.635 &      15.580 &       0.325  &    -5.74  &    -6.07 \\
  7  &     7952  &   0.0792 & $<$ 0.01  &      59.063 &      30.928 &       0.139  &     8.23  &     9.55 \\
  8  &     7951  &  -0.0234 &     1.85  &       1.813 &      39.772 &       0.104  &     0.19  &     0.19 \\
  9  &     7905  &  -0.0544 & $<$ 0.01  &    -135.524 &      98.548 &       0.040  &    -5.43  &    -5.44 \\
 10  &     7952  &   0.0112 &    15.98  &      41.186 &      81.962 &       0.051  &     2.09  &     2.10 \\
 11  &     7951  &   0.0421 &     0.01  &      77.948 &      58.581 &       0.070  &     5.43  &     5.21 \\
 12  &     7905  &   0.0843 & $<$ 0.01  &     125.368 &      59.849 &       0.065  &     8.17  &     8.09 \\
 13  &     7951  &  -0.0337 &     0.13  &    -103.639 &      89.066 &       0.048  &    -4.94  &    -3.91 \\
 14  &     7949  &   0.0132 &    11.94  &      16.996 &      74.317 &       0.058  &     0.99  &     0.87 \\
 15  &     7899  &  -0.0003 &    49.10  &      16.916 &      70.642 &       0.059  &     1.00  &     0.98 \\
 16  &     7947  &   0.0219 &     2.55  &       9.961 &      70.057 &       0.065  &     0.65  &     0.55 \\
 17  &     7944  &  -0.0096 &    19.57  &      -8.173 &      34.141 &       0.129  &    -1.06  &    -1.02 \\
 18  &     7821  &   0.0223 &     2.42  &       3.532 &      14.848 &       0.289  &     1.02  &     0.87 \\
 19  &     3211  &   0.1063 & $<$ 0.01  &      23.486 &      12.282 &       0.498  &    11.70  &    10.38 \\
 21  &     4134  &  -0.0747 & $<$ 0.01  &      -7.967 &       7.640 &       0.701  &    -5.59  &    -5.78 \\
 22  &     7952  &  -0.0149 &     9.17  &      -0.977 &       8.226 &       0.450  &    -0.44  &    -0.49 \\
 23  &     7905  &   0.0390 &     0.03  &       7.824 &      11.336 &       0.340  &     2.66  &     3.12
\enddata
\end{deluxetable}

\clearpage

%Table: Q_BEAST and Dust coefficient -----------------------------------------
\begin{deluxetable}{ccccccccc}
\tabletypesize{\scriptsize}
\tablecaption{Correlation results between BEAST Q-band map and
dust template}%. \label{tbl-1}}
\tablewidth{0pt}
\tablehead{
\colhead{Section} & 
\colhead{Number of} & 
\colhead{Correlation}   &
\colhead{Probability} &
\colhead{$\tilde\alpha$}  & 
\colhead{$\sigma_{\tilde\alpha}$} & 
\colhead{$\sigma_{Gal}$} &
\colhead{$\Delta T \equiv \tilde\alpha \cdot\sigma_{Gal}$}     & 
\colhead{$\Delta T / \sigma_B$} \\
\colhead{} &
\colhead{pixels} &
\colhead{Coeff.} &
\colhead{[$\%$]} &
\colhead{[$\mu${K/(MJy/sr)}]}  & 
\colhead{[$\mu${K/(MJy/sr)}]} & 
\colhead{[$MJy/sr$]} & 
\colhead{[$\mu$K]} & 
\colhead{[$\%$]}
}
\startdata
  0  &     7866  &  -0.0143 &    10.21  &      -1.108 &       5.800 &       0.308  &    -0.34  &    -0.75 \\
  1  &     7866  &  -0.1378 & $<$ 0.01  &     -33.899 &      12.862 &       0.174  &    -5.90  &   -12.25 \\
  2  &     7866  &  -0.0010 &    46.51  &      -2.607 &      11.030 &       0.222  &    -0.58  &    -1.21 \\
  3  &     7723  &   0.4429 & $<$ 0.01  &      14.550 &       1.183 &       2.034  &    29.60  &    47.02 \\
  4  &     2059  &   0.6713 & $<$ 0.01  &      23.965 &       1.074 &       4.224  &   101.22  &    67.48 \\
  6  &     5361  &   0.0149 &    13.72  &       3.324 &       8.136 &       0.348  &     1.16  &     2.48 \\
  7  &     7866  &   0.1098 & $<$ 0.01  &      39.823 &      17.641 &       0.144  &     5.73  &    11.75 \\
  8  &     7866  &  -0.0469 & $<$ 0.01  &      -9.518 &      22.134 &       0.108  &    -1.03  &    -2.20 \\
  9  &     7820  &   0.0602 & $<$ 0.01  &      60.040 &      54.567 &       0.041  &     2.46  &     4.78 \\
 10  &     7866  &   0.0502 & $<$ 0.01  &      79.322 &      46.521 &       0.054  &     4.27  &     7.42 \\
 11  &     7866  &  -0.1068 & $<$ 0.01  &     -91.406 &      33.964 &       0.069  &    -6.30  &   -14.33 \\
 12  &     7820  &   0.0491 & $<$ 0.01  &      27.629 &      35.994 &       0.065  &     1.80  &     2.88 \\
 13  &     7866  &  -0.0949 & $<$ 0.01  &    -139.652 &      49.966 &       0.049  &    -6.88  &   -12.00 \\
 14  &     7866  &   0.1019 & $<$ 0.01  &      14.798 &      39.502 &       0.062  &     0.92  &     1.41 \\
 15  &     7820  &  -0.0165 &     7.19  &     -23.783 &      40.560 &       0.061  &    -1.45  &    -2.32 \\
 16  &     7866  &   0.1248 & $<$ 0.01  &     121.100 &      41.298 &       0.068  &     8.21  &    10.63 \\
 17  &     7866  &  -0.0393 &     0.02  &     -37.852 &      19.775 &       0.134  &    -5.07  &    -7.33 \\
 18  &     7820  &  -0.1362 & $<$ 0.01  &     -19.961 &       7.869 &       0.293  &    -5.86  &   -11.28 \\
 19  &     3135  &  -0.1578 & $<$ 0.01  &     -19.227 &       7.309 &       0.497  &    -9.55  &   -16.93 \\
 21  &     4187  &  -0.0319 &     1.94  &       0.070 &       4.039 &       0.731  &     0.05  &     0.12 \\
 22  &     7866  &  -0.0297 &     0.42  &       2.066 &       4.462 &       0.458  &     0.95  &     2.00 \\
 23  &     7820  &  -0.0121 &    14.17  &       1.747 &       6.321 &       0.371  &     0.65  &     1.35 \\
\enddata

\end{deluxetable}

\clearpage

\begin{deluxetable}{cccccccc}
\tabletypesize{\scriptsize}
\tablecaption{Correlation results between BEAST Ka-band map and
foreground templates after different cuts in galactic latitude}
\tablewidth{0pt}
\tablehead{
\colhead{Galactic cut} &
\colhead{Number of} &
\colhead{$\tilde\alpha_{H\alpha}$} &
\colhead{$\Delta T_{H\alpha}/ \sigma_{B}$} & 
\colhead{$\tilde\alpha_{sync}$} & 
\colhead{$\Delta T_{sync}/ \sigma_{B}$} & 
\colhead{$\tilde\alpha_{dust}$} & 
\colhead{$\Delta T_{dust}/ \sigma_{B}$} \\
\colhead{$\mid$b$\mid$ [$^{\circ}$]} & 
\colhead{pixels} & 
\colhead{[$\mu${K/R}]} & 
\colhead{[$\%$]} & 
\colhead{[$\mu${K/K}]} & 
\colhead{[$\%$]} & 
\colhead{[$\mu${K/(MJy/sr)}]} & 
\colhead{[$\%$]}
}

\startdata
  2.50  &  182349    &  5.5 $\pm$   0.2    &   14.11    &  4.2 $\pm$ 0.1    &   34.61 &    11.9 $\pm$   0.3  &    25.53 \\
  7.50  &  165947    &  6.9 $\pm$   0.3    &   19.99    &  1.9 $\pm$ 0.7    &    2.16 &    10.9 $\pm$   0.6  &    13.46 \\
 12.50  &  148942    &  8.3 $\pm$   0.4    &   15.89    & 14.0 $\pm$ 5.1    &    2.44 &    35.0 $\pm$   2.2  &    13.86 \\
 17.50  &  130519    & -4.3 $\pm$   2.6    &   -1.57    &  8.2 $\pm$ 5.8    &    1.37 &     4.4 $\pm$   4.7  &     0.88 \\
 22.50  &  106959    & -5.0 $\pm$  12.0    &   -0.45    & 10.0 $\pm$ 6.7    &    1.63 &     9.5 $\pm$   8.7  &     1.2
\enddata

\end{deluxetable}

%Table: Cutting in latitude Q_BEAST and Halpha coefficient -------------------------------------

%\begin{table}
%\caption{Correlation results between BEAST Q-band map and
%foreground templates after different cuts in galactic latitude}
%\[
%\begin{array}{cccccccc}
%\hline\hline \noalign{\smallskip}  Galactic cut & 
%Number of & \tilde\alpha_{H\alpha} & \Delta T_{H\alpha}/ \sigma_{B} &
%\tilde\alpha_{sync} & \Delta T_{sync}/ \sigma_{B} & \tilde\alpha_{dust} & \Delta T_{dust}/ \sigma_{B} \\
% $\mid$b$\mid$ [$^{\circ] &  pixels  &
%[\mu${K/R] & [${\%] & [\mu${K/K] & [${\%] & [\mu${K/(MJy/sr)] & [${\%] \\
% \hline \noalign{\smallskip}

\begin{deluxetable}{cccccccc}
\tabletypesize{\scriptsize}
\tablecaption{Correlation results between BEAST Q-band map and
foreground templates after different cuts in galactic latitude}
\tablewidth{0pt}
\tablehead{
\colhead{Galactic cut} &
\colhead{Number of} &
\colhead{$\tilde\alpha_{H\alpha}$} &
\colhead{$\Delta T_{H\alpha}/ \sigma_{B}$} & 
\colhead{$\tilde\alpha_{sync}$} & 
\colhead{$\Delta T_{sync}/ \sigma_{B}$} & 
\colhead{$\tilde\alpha_{dust}$} & 
\colhead{$\Delta T_{dust}/ \sigma_{B}$} \\
\colhead{$\mid$b$\mid$ [$^{\circ]}$} & 
\colhead{pixels} & 
\colhead{[$\mu${K/R}]} & 
\colhead{[$\%$]} & 
\colhead{[$\mu${K/K}]} & 
\colhead{[$\%$]} & 
\colhead{[$\mu${K/(MJy/sr)}]} & 
\colhead{[$\%$]}
}
\startdata

   2.50  &  180412 &    2.4 $\pm$  0.1 &   12.47 &  1.73 $\pm$   0.04 &   27.42 &   7.0 $\pm$   0.2 &   29.45 \\
   7.50  &  164235 &    3.4 $\pm$  0.2 &   20.02 &  1.55 $\pm$   0.32 &    3.42 &   5.1 $\pm$   0.4 &   11.99 \\
  12.50  &  147474 &    4.2 $\pm$  0.2 &   16.14 & -3.7  $\pm$   2.9  &   -1.24 &  16.2 $\pm$   1.2 &   12.65 \\
  17.50  &  129353 &   -3.3 $\pm$  1.5 &   -2.35 & -3.0  $\pm$   3.2  &   -0.95 &  -3.7 $\pm$   2.6 &   -1.46 \\
  22.50  &  106350 &  -15.0 $\pm$  6.9 &   -2.47 &  1.6  $\pm$   3.8  &    0.46 &  -3.9 $\pm$   4.9 &   -0.90 \\

\enddata

\end{deluxetable}

\end{document}